\documentclass{aa}

\usepackage{graphicx}
\usepackage{float}
\usepackage{color}
\usepackage{amsmath}
\usepackage[colorlinks=true, linkcolor=blue, citecolor=blue, urlcolor=blue]{hyperref}

\usepackage{natbib}
\bibpunct{(}{)}{;}{a}{}{,}

\usepackage{txfonts}

\makeatletter
\let\linenumbers\@gobble
\let\endlinenumbers\@empty
\makeatother

\begin{document}

    \title{The ALMA-QUARKS survey: Extensive detection of acetamide in multiple high-mass star-forming regions
                      \thanks{Corresponding authors: \\
                      Xuefang Xu, \email{xuefang\_xu@cqu.edu.cn}\\
                      Qian Gou, \email{qian.gou@cqu.edu.cn}\\
                      Tie Liu, \email{liutie@shao.ac.cn}}}

   \subtitle{}

   \author{Chunguo Duan\inst{1,2},
          Xuefang Xu\inst{1,2},    
          Qian Gou\inst{1,2},  
          Tie Liu\inst{3,4},               
          Laurent Pagani\inst{5},                       
          Fengwei Xu\inst{6,7},       
          Ke Wang\inst{7},
          Xunchuan Liu\inst{3,4,8},
          Jun Kang\inst{1,2}
          Mingwei He\inst{1,2}
          Jiaxiang Jiao\inst{1,2}
          }

   \institute{
            \inst{1} School of Chemistry and Chemical Engineering, Chongqing University, Daxuecheng South Rd. 55, Chongqing 401331, People’s Republic of China\\
            \inst{2} Chongqing Key Laboratory of Chemical Theory and Mechanism, Chongqing University, Daxuecheng South Rd. 55, Chongqing 401331, People’s Republic of China\\
            \inst{3} Shanghai Astronomical Observatory, Chinese Academy of Sciences, Nandan Rd. 80, Shanghai 200030, People’s Republic of China\\
            \inst{4}  Key Laboratory for Research in Galaxies and Cosmology, Shanghai Astronomical Observatory, Chinese Academy of Sciences, Nandan Rd. 80, Shanghai 200030, People’s Republic of China\\
            \inst{5} LUX, Observatoire de Paris, PSL Research University, CNRS, Sorbonne Universités, 75014 Paris, France\\                                      
            \inst{6} Department of Astronomy, Peking University, 5 Yiheyuan Road, Haidian District, Beijing 100871, People’s Republic of China\\
            \inst{7} Kavli Institute for Astronomy and Astrophysics, Peking University, 5 Yiheyuan Road, Haidian District, Beijing 100871, People’s Republic of China\\                                
           \inst{8} Leiden Observatory, Leiden University, P.O. Box 9513, 2300RA Leiden, The Netherlands\\                         
           }

  \abstract
   {Acetamide (CH$_{3}$CONH$_{2}$), a key interstellar amide and a methyl derivative of formamide (NH$_{2}$CHO), has been sparsely detected, limiting insights into its prebiotic relevance. We present the first systematic survey for acetamide toward 52 hot molecular cores using ALMA Band 6 data. Acetamide has been detected in 10 cores, markedly expanding the inventory of known emitters. The derived column densities of acetamide range from $(2.5\pm0.9)\times10^{14}$ to $(1.5\pm0.6)\times10^{16}$ cm$^{-2}$, compared to formamide’s $(1.1\pm0.1)\times10^{15}$ to $(6.9\pm0.4)\times10^{16}$ cm$^{-2}$. The nearly constant abundance ratios ($\sim$3--9) and strong abundance correlation between the two amides across sources suggest a chemically linked formation pathway, likely on grain surfaces. The presence of peptide-like molecules in these regions implies that complex organic species can survive star formation processes, offering a potential pathway toward prebiotic chemistry. These findings constrain the dominant grain-surface formation routes of acetamide, confirm its broader prevalence in high-mass star-forming regions, and underscore the importance of targeted amide surveys in tracing the chemical evolution toward prebiotic complexity.}


\titlerunning{Acetamide} \authorrunning{Duan et al.}\maketitle

\section{Introduction} \label{sec:sec1}

Life on Earth appeared about 700 million years after planetary formation \citep[$\sim$3.8 billion years ago;][]{Pea18}, yet the mechanisms driving its origin remain unknown. What is clear is that life depends on recurring chemical motifs. Among them, the amide functional group (–NH–C(O)–) is central to peptide bonds and essential for biochemistry \citep{Pau51}.

Understanding how amides appeared on early Earth is vital for tracing the chemical origin of life. While they may have formed through prebiotic processes on Earth \citep{Pat15}, an alternative scenario involves their formation in space and subsequent delivery by meteorites or comets \citep{Chy90, Chy92}. Such molecules could have originated in the interstellar medium (ISM), within the parental molecular cloud of the Solar System. This raises the possibility that analogous prebiotic pathways operate in other planetary systems across the Milky Way.

\begin{table*}[htp]
	\caption{Parameters of HMCs sample.}\label{table:table 1}
	\centering
	\footnotesize
	\renewcommand{\arraystretch}{1}
	\begin{tabular}{ccccccc}    
	    \hline
	    \hline
ID & Sources name & R.A. & decl. & $V_{\rm LSR}$$^{\rm a}$ & $D_{\rm GC}$$^{\rm b}$ & $^{12}$C/$^{13}$C$^{\rm c}$ \\          
      {} & {} & {(J2000)} & {(J2000)} & {(km s$^{-1}$)} & {(kpc)} & {} \\                     
    	\hline                                                                          
1	  &  I08303	      &  08:32:09.190	 &  -43:13:44.30	&   14.5	  &  9.0 	&  63.7  \\
2	  &  I08470	      &  08:48:47.720	 &  -42:54:22.00	&   12.1	  &  8.8 	&  62.7  \\
3	  &  I09018	      &  09:03:32.820	 &  -48:28:06.20	&    9.9	  &  8.8 	&  62.7  \\
4	  &  I11298	      &  11:32:06.030	 &  -62:12:20.50	&   33.4	  & 10.1 	&  68.9  \\
5	  &  I12326	      &  12:35:34.810	 &  -63:02:32.10	&  -39.5	  &  7.2 	&  55.1  \\
6	  &  I13079	      &  13:11:13.730	 &  -62:34:40.20	&  -41.3	  &  6.9 	&  53.7  \\
7	  &  I13134	      &  13:16:42.990	 &  -62:58:29.30	&  -32.0	  &  6.9 	&  53.7  \\
8	  &  I13140	      &  13:17:15.900	 &  -62:42:27.00	&  -34.4	  &  6.9 	&  53.7  \\
9	  &  I13471	      &  13:50:42.100	 &  -61:35:14.90	&  -57.8	  &  6.4 	&  51.3  \\
10	&  I13484	      &  13:51:58.640	 &  -61:15:43.30	&  -55.3	  &  6.4 	&  51.3  \\
11	&  I14498	      &  14:53:42.530	 &  -59:08:53.20	&  -50.2	  &  6.4 	&  51.3  \\
12	&  I15254	      &  15:29:19.480	 &  -56:31:23.20	&  -68.6	  &  5.7 	&  47.9  \\
13	&  I15437	      &  15:47:33.110	 &  -53:52:43.90	&  -83.4	  &  5.0 	&  44.6  \\
14	&  I15520	      &  15:55:48.390	 &  -52:43:09.80	&  -41.8	  &  6.2 	&  50.3  \\
15	&  I16060	      &  16:09:52.850	 &  -51:54:54.70	&  -92.1	  &  4.5 	&  42.2  \\
16	&  I16065	      &  16:10:19.600	 &  -52:06:07.10	&  -62.5	  &  5.2 	&  45.6  \\
17	&  I16071	      &  16:10:59.010	 &  -51:50:21.60	&  -86.5	  &  4.5 	&  42.2  \\
18	&  I16076	      &  16:11:27.120	 &  -51:41:56.90	&  -87.8	  &  4.5 	&  42.2  \\
19	&  I16164	      &  16:20:10.910	 &  -50:53:15.50	&  -56.7	  &  5.4 	&  46.5  \\
20	&  I16172	      &  16:21:02.930	 &  -50:35:11.60	&  -53.3	  &  5.4 	&  46.5  \\
21	&  I16272	      &  16:30:58.020	 &  -48:43:46.60	&  -46.8	  &  5.8 	&  48.4  \\
22	&  I16318	      &  16:35:33.200	 &  -47:31:11.30	& -120.8	  &  3.3 	&  36.5  \\
23	&  I16344	      &  16:38:10.380	 &  -47:04:56.70	&  -49.7	  &  5.4 	&  46.5  \\
24	&  I16348	      &  16:38:29.420	 &  -47:00:39.70	&  -47.8	  &  5.4 	&  46.5  \\
25	&  I16351	      &  16:38:50.610	 &  -47:27:59.70	&  -40.8	  &  5.7 	&  47.9  \\
26	&  I16458	      &  16:49:30.410	 &  -45:17:53.60	&  -50.9	  &  5.1 	&  45.1  \\
27	&  I16484	      &  16:52:03.990	 &  -46:08:24.60	&  -32.4	  &  6.4 	&  51.3  \\
28	&  I16547	      &  16:58:17.260	 &  -42:52:04.50	&  -30.6	  &  5.8 	&  48.4  \\
29	&  I17008	      &  17:04:23.200	 &  -40:44:24.90	&  -17.3	  &  6.1 	&  49.9  \\
30	&  I17016	      &  17:05:11.020	 &  -41:29:07.80	&  -26.8	  &  7.0 	&  54.2  \\
31	&  I17158	      &  17:19:20.340	 &  -39:03:53.30	&  -16.6	  &  5.1 	&  45.1  \\
32	&  I17175 MM1	  &  17:20:53.570	 &  -35:46:59.80	&   -8.7	  &  7.0 	&  54.2  \\
33	&  I17175 MM2	  &  17:20:53.570	 &  -35:46:59.80	&   -8.7	  &  7.0 	&  54.2  \\
34	&  I17220	      &  17:25:24.990	 &  -36:12:41.10	&  -97.2	  &  1.3 	&  27.0  \\
35	&  I17233	      &  17:26:42.800	 &  -36:09:16.80	&   -3.1	  &  7.0 	&  54.2  \\
36	&  I17441	      &  17:47:19.790	 &  -28:23:05.70	&   50.8	  &  0.2 	&  21.7  \\
37	&  I18032	      &  18:06:14.300	 &  -20:31:35.00	&    4.4	  &  3.4 	&  37.0  \\
38	&  I18056	      &  18:08:38.180	 &  -19:51:49.00	&   66.4	  &  1.6 	&  28.4  \\
39	&  I18089	      &  18:11:51.060	 &  -17:31:27.20	&   32.9	  &  5.9 	&  48.9  \\
40	&  I18117	      &  18:14:39.250	 &  -17:51:59.80	&   37.1	  &  5.9 	&  48.9  \\
41	&  I18159	      &  18:18:54.340	 &  -16:47:45.90	&   22.4	  &  6.9 	&  53.7  \\
42	&  I18182	      &  18:21:09.220	 &  -14:31:46.80	&   59.3	  &  4.1 	&  40.3  \\
43	&  I18236	      &  18:26:25.650	 &  -12:03:57.60	&   26.1	  &  6.3 	&  50.8  \\
44	&  I18290	      &  18:31:42.980	 &  -09:22:26.00	&   84.0	  &  4.0 	&  39.8  \\
45	&  I18316	      &  18:34:20.580	 &  -05:59:41.60	&   42.7	  &  6.5 	&  51.8  \\
46	&  I18411	      &  18:43:46.260	 &  -03:35:23.90	&  103.5	  &  4.0 	&  39.8  \\
47	&  I18469	      &  18:49:33.150	 &  -01:29:06.20	&   86.6	  &  4.7 	&  43.2  \\
48	&  I18507+0110	&  18:53:18.120	 &  +01:15:00.10	&   58.2	  &  7.1 	&  54.6  \\
49	&  I18507+0121	&  18:53:18.150	 &  +01:25:22.40	&   57.8	  &  7.1 	&  54.6  \\
50	&  I18517	      &  18:54:14.110	 &  +04:41:43.10	&   43.9	  &  6.6 	&  52.2  \\
51	&  I19078	      &  19:10:13.410	 &  +09:06:10.40	&    6.2	  &  7.6 	&  57.0  \\
52	&  I19095	      &  19:11:53.900	 &  +09:35:45.90	&   43.8	  &  5.8 	&  48.4  \\
	    \hline
	    \hline					
	\end{tabular}
	\tablefoot{\\
       $^{\rm a}$ The V$_{\rm LSR}$ was retrieved from \citet{Liu24}.\\
       $^{\rm b}$ The D$_{\rm GC}$ was retrieved from \citet{Liu20}.\\
       $^{\rm c}$ The $^{12}$C/$^{13}$C ratio was derived from the formula proposed by \citet{Yan23}.}
\end{table*}

To assess the prebiotic chemistry relevance of amides, it is crucial to determine their presence in star- and planet-forming regions. Formamide (NH$_{2}$CHO), the simplest amide and a key prebiotic precursor, has been widely detected across various astrophysical environments (e.g., \citealt{Zhe24}; \citealt{Zen23}; \citealt{Liu22}; \citealt{Lig20}, and the review by \citealt{Lop19} for a comprehensive list of observations). Recent observations further confirmed its prevalence in star forming regions \citep{Xu25}. In contrast, acetamide (CH$_{3}$CONH$_{2}$), a methylated derivative of formamide representing a potential step toward more complex biomolecules, has been detected in only a handful of star-forming regions, including Sgr B2 \citep{Hol06, Hal11, Bel17, Zhe24}, NGC 6334I \citep{Lig20}, G31.41+0.31 \citep{Col21}, and Orion KL \citep{Cer16}, as well as the intermediate-mass protostar Serpens SMM1-a \citep{Lig22} and the low-mass protostar IRAS 16293-2422B \citep{Lig17}. Notably, no large-sample survey has yet confirmed widespread acetamide emission.

\begin{figure*}[!htp]
	\centering
    \includegraphics[width=0.95\textwidth]{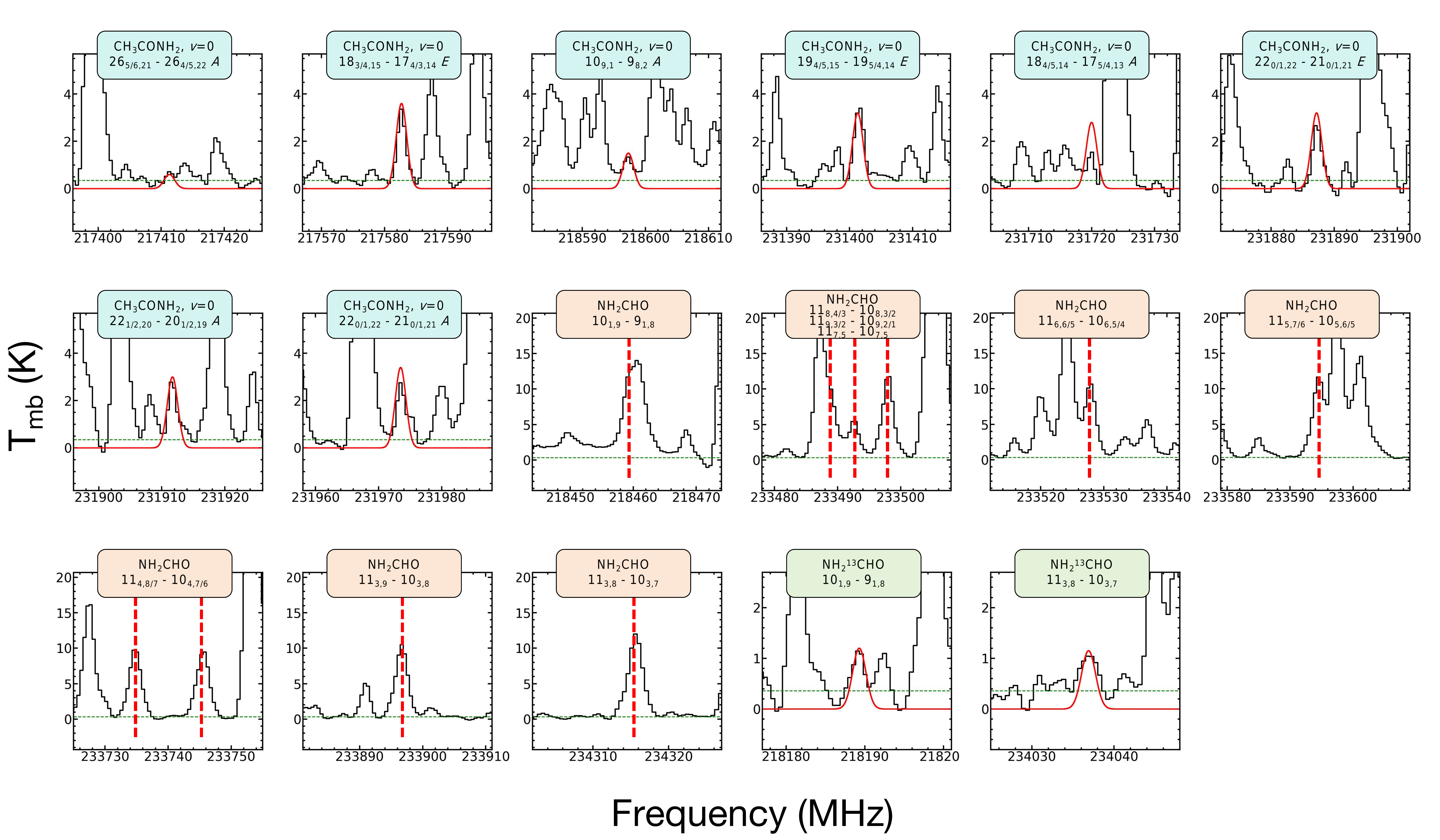}
	\caption{Observed (black) and modeled (red) spectra of CH$_{3}$CONH$_{2}$, NH$_{2}$CHO, and NH$_{2}$$^{13}$CHO in I17175 MM2. Clearly resolved or slightly blended transitions are shown. Green dashed lines mark the 3$\sigma$ noise level. No modeled spectrum is shown for NH$_{2}$CHO due to optical thickness. Red dashed lines indicate the rest frequencies of NH$_{2}$CHO transitions. Spectra for other sources are shown in Fig. \ref{fig:fig. B1}.
	\label{fig:fig. 1}} 
\end{figure*}

\begin{figure*}[!htp]
	\centering
    \includegraphics[width=0.95\textwidth]{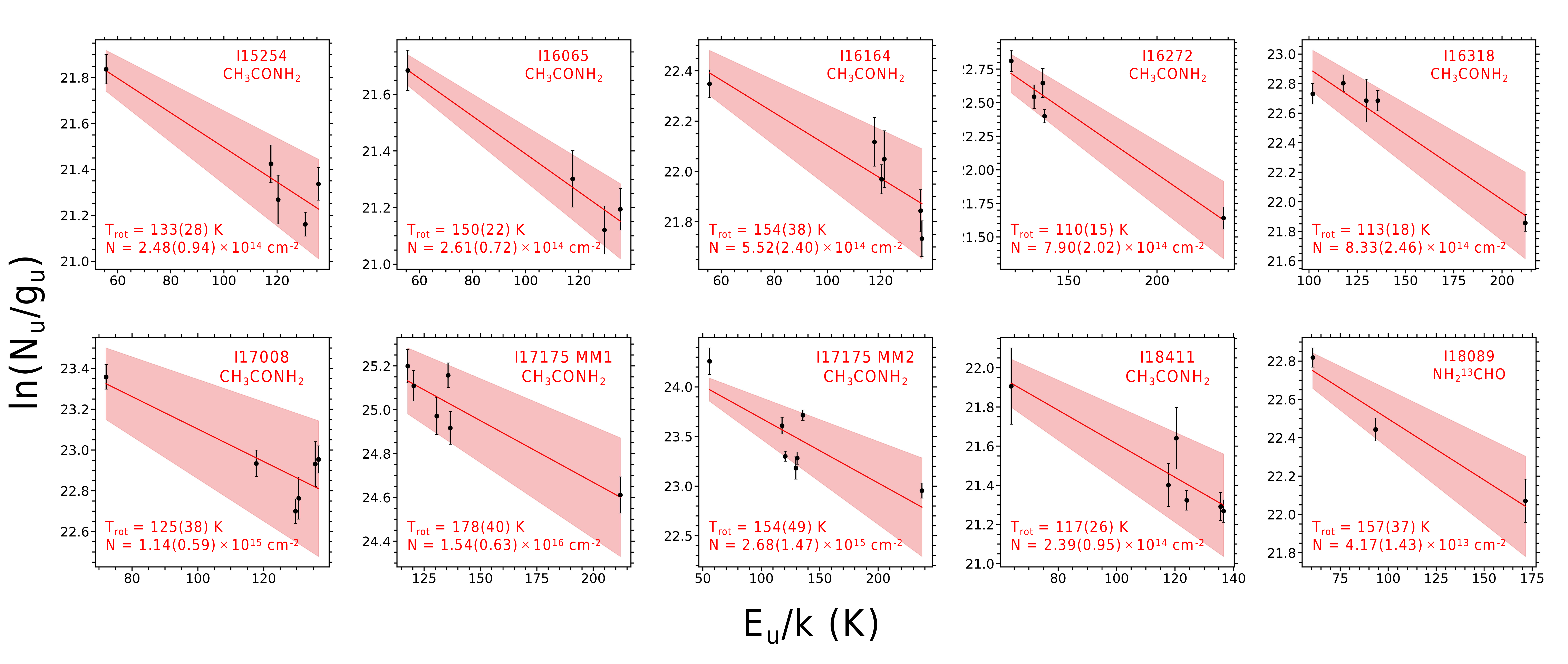}
	\caption{Rotational diagrams of CH$_{3}$CONH$_{2}$ and NH$_{2}$$^{13}$CHO. Black points represent the observed data, while red lines indicate the best-fit results. Shaded red regions denote the associated fitting uncertainties. Each panel includes the source name and molecule in the top-right corner, with the derived rotational temperature ($T_{\rm rot}$) and column density ($N$) shown in the bottom-left.
	\label{fig:fig. 2}} 
\end{figure*}

Expanding the number of amide detections is essential in order to evaluate its role in prebiotic chemistry. Although astrochemical models have incorporated potential reaction networks for formamide and acetamide \citep{Bel17, Bel19, Gar22}, observational constraints -- especially for larger amides -- remain sparse. To address this gap, we present the first systematic search for acetamide across 52 hot molecular cores (HMCs) using Band 6 observations from the Atacama Large Millimeter/submillimeter Array (ALMA). Our detections substantially increase the number of known acetamide sources, enabling a comparative analysis with formamide. These findings offer new insights into interstellar amide chemistry and its potential connection to prebiotic molecular evolution. 

This paper is structured as follows. Sect. \ref{sec:sec2} outlines the source sample and the observational setup. Sect. \ref{sec:sec3} presents detection results and derivation of column densities, excitation temperatures. Sect. \ref{sec:sec4} discusses chemical correlations and formation pathways. Sect. \ref{sec:sec5} summarizes our conclusions.

\begin{table*}[htp]
	\caption{Parameters of detected amide molecules.}\label{table:table 2}
	\centering
	\fontsize{4.5}{7}\selectfont
	\renewcommand{\arraystretch}{1.3}
	\begin{tabular}{lcccccccccccccccc}    
	    \hline
	    \hline
      Source & $\theta_{\rm source}$ & \multicolumn{4}{c}{NH$_{2}$$^{13}$CHO} & \multicolumn{4}{c}{NH$_{2}$CHO} & \multicolumn{5}{c}{CH$_{3}$CONH$_{2}$} & Ratio$^{\mathrm{c}}$ \\                                                                                                       
\cline{3-6} \cline{7-10} \cline{11-15}                                                                                                                                                                                                                                                               
&  & FTMW & $V_{\rm LSR}$ & $T_{\mathrm{ex}}$  & $N$  & FTMW & $V_{\rm LSR}$ & $N$  & $X$${_{\rm CH_3OH}}$$^{\mathrm{b}}$ & FTMW & $V_{\rm LSR}$ & $T_{\mathrm{ex}}$ & $N$ & $X$${_{\rm CH_3OH}}$$^{\mathrm{b}}$ & {} & \\                                                                           
& (arcsec) &  (km $\rm s^{-1})$ & (km $\rm s^{-1}$) & (K) & (cm$^{-2}$) & (km $\rm s^{-1}$) & (km $\rm s^{-1}$) & (cm$^{-2}$) &  & (km $\rm s^{-1}$) & (km $\rm s^{-1}$) & (K) & (cm$^{-2}$) &  & {} & \\                                                                                            
\hline                                                                                                                                                                                                                                                                                               
I15254     & 1.0 & 2.6 & -66.6 & 133$^{\mathrm{a}}$ & (2.4$\pm$0.2)$\times$10$^{13}$    & 2.7 & -66.6 &  (1.1$\pm$0.1)$\times$10$^{15}$  & (6.3$\pm$0.7)$\times$10$^{-4}$ & 3.4 & -66.6 & 133$\pm$28          & (2.5$\pm$0.9)$\times$10$^{14}$ & (1.4$\pm$0.5)$\times$10$^{-4}$ & 4.6$\pm$1.8 \\     
I16065     & 2.9 & 2.7 & -60.1 & 150$^{\mathrm{a}}$ & (4.2$\pm$0.3)$\times$10$^{13}$    & 5.4 & -59.5 &  (1.9$\pm$0.1)$\times$10$^{15}$  & (1.5$\pm$0.2)$\times$10$^{-3}$ & 3.1 & -59.5 & 150$\pm$22          & (2.6$\pm$0.7)$\times$10$^{14}$ & (2.0$\pm$0.6)$\times$10$^{-4}$ & 7.3$\pm$2.1 \\     
I16164     & 3.8 & 3.3 & -59.7 & 157$^{\mathrm{a}}$ & (7.4$\pm$0.5)$\times$10$^{13}$    & 5.3 & -59.7 &  (3.5$\pm$0.2)$\times$10$^{15}$  & (3.8$\pm$0.3)$\times$10$^{-3}$ & 2.8 & -59.7 & 154$\pm$38          & (5.5$\pm$2.4)$\times$10$^{14}$ & (6.1$\pm$2.6)$\times$10$^{-4}$ & 6.2$\pm$2.7 \\     
I16272     & 1.4 & 2.8 & -45.9 & 110$^{\mathrm{a}}$ & (1.2$\pm$0.2)$\times$10$^{14}$    & 5.1 & -46.3 &  (5.6$\pm$1.2)$\times$10$^{15}$  & (1.8$\pm$0.4)$\times$10$^{-3}$ & 2.7 & -46.3 & 110$\pm$15          & (7.9$\pm$2.0)$\times$10$^{14}$ & (2.5$\pm$0.6)$\times$10$^{-4}$ & 7.1$\pm$2.3 \\     
I16318     & 2.3 & 4.7 & -118.2 & 113$^{\mathrm{a}}$ & (2.0$\pm$0.3)$\times$10$^{14}$    & 7.7 & -118.5 &  (7.4$\pm$1.0)$\times$10$^{15}$  & (2.7$\pm$0.4)$\times$10$^{-3}$ & 3.6 & -117.8 & 113$\pm$18          & (8.3$\pm$2.5)$\times$10$^{14}$ & (3.0$\pm$0.9)$\times$10$^{-4}$ & 8.9$\pm$2.9 \\  
I17008     & 1.3 & 4.4 & -19.5 & 125$^{\mathrm{a}}$ & (1.4$\pm$0.3)$\times$10$^{14}$    & 7.4 & -19.1 &  (7.0$\pm$1.7)$\times$10$^{15}$  & (2.2$\pm$0.5)$\times$10$^{-3}$ & 4.8 & -19.8 & 125$\pm$38          & (1.1$\pm$0.6)$\times$10$^{15}$ & (3.6$\pm$1.9)$\times$10$^{-4}$ & 6.1$\pm$3.5 \\     
I17175 MM1 & 2.1 & 4.1 & -5.1 & 178$^{\mathrm{a}}$ & (1.3$\pm$0.1)$\times$10$^{15}$    & 7.3 & -4.9 &  (6.9$\pm$0.4)$\times$10$^{16}$  & (1.1$\pm$0.1)$\times$10$^{-2}$ & 4.4 & -4.5 & 178$\pm$40          & (1.5$\pm$0.6)$\times$10$^{16}$ & (2.4$\pm$1.0)$\times$10$^{-3}$ & 4.5$\pm$1.9 \\        
I17175 MM2 & 1.6 & 3.8 & -8.7 & 154$^{\mathrm{a}}$ & (1.51$\pm$0.01)$\times$10$^{14}$  & 3.1 & -8.7 &  (8.2$\pm$0.7)$\times$10$^{15}$  & (1.3$\pm$0.1)$\times$10$^{-3}$ & 2.5 & -8.7 & 154$\pm$49          & (2.7$\pm$1.5)$\times$10$^{15}$ & (4.3$\pm$2.3)$\times$10$^{-4}$ & 3.0$\pm$1.7 \\        
I18089     & 1.8 & 1.9 & 29.9 & 157$\pm$37         & (4.2$\pm$1.4)$\times$10$^{13}$    & 4.4 & 31.4 &  (2.0$\pm$0.7)$\times$10$^{15}$  & (7.9$\pm$2.7)$\times$10$^{-4}$ & 4.1 & 29.9 & 157$^{\mathrm{a}}$  & (6.9$\pm$1.3)$\times$10$^{14}$ & (2.7$\pm$0.5)$\times$10$^{-4}$ & 2.9$\pm$1.2 \\        
I18411     & 0.9 & 2.1 & 105.5 & 117$^{\mathrm{a}}$ & (3.0$\pm$0.3)$\times$10$^{13}$    & 5.2 & 105.8 &  (1.2$\pm$0.1)$\times$10$^{15}$  & (1.3$\pm$0.1)$\times$10$^{-3}$ & 2.2 & 105.5 & 117$\pm$26          & (2.4$\pm$1.0)$\times$10$^{14}$ & (2.6$\pm$1.1)$\times$10$^{-4}$ & 4.9$\pm$2.0 \\         
	    \hline
	    \hline					
	\end{tabular}
	\tablefoot{\\
       $^{\rm a}$ For NH$_{2}$$^{13}$CHO, excitation temperature fitting errors are not provided due to fewer than three clean transitions. Column densities are derived by assuming the excitation temperatures for CH$_{3}$CONH$_{2}$. For CH$_{3}$CONH$_{2}$ in I18089, the column density is derived by assuming fixed excitation temperatures of NH$_{2}$$^{13}$CHO, due to the small energy range of the upper level for the three detected lines of CH$_{3}$CONH$_{2}$.\\
       $^{\rm b}$ Molecular abundances relative to CH$_{3}$OH, where CH$_{3}$OH column densities are adopted from \citet{Qin22}.\\
       $^{\rm c}$ Molecular ratios of NH$_{2}$CHO to CH$_{3}$CONH$_{2}$.}
\end{table*}

\section{Observations} \label{sec:sec2}

This study is based on a sample of HMCs summarized in Table \ref{table:table 1}. The targets were selected from the QUARKS survey \citep[Querying Underlying mechanisms of massive star formation with ALMA-Resolved gas Kinematics and Structures;] []{Liu24}, based on HMC candidates previously identified by \citet{Qin22}. QUARKS is a 1.3 mm follow-up to the ALMA Three-millimeter Observations of Massive Star-forming Regions \citep[ATOMS;] []{Liu20}, aiming to resolve substructures within 3 mm core clusters in massive star-forming clumps \citep{Liu20}. 

The QUARKS survey was conducted with ACA and 12 m arrays at Band 6 towards 139 protoclusters from October 2021 to June 2024. Further details on the observational setup and data reduction are provided in \citet{Liu24}. Briefly, The ACA and TM2 combined dataset of this work yields an angular resolution of $\sim$1.3$^{\prime \prime}$, and a typical rms noise level of $\sim$5 mJy beam$^{-1}$. The flux calibration uncertainty is estimated at $\sim$10\%. Four spectral windows (SPW 1-4), each with a bandwidth of 1.875 GHz, were centered at 217.918, 220.319, 231.370, and 233.520 GHz. The uniform spectral resolution of 976.56 kHz corresponds to a velocity resolution ($\delta$$V$) of 1.2 km s$^{-1}$, sufficient to spectrally resolve broad molecular lines.

\section{Results} \label{sec:sec3}

\subsection{Molecular Line Identifications} \label{sec:sec3.1}

Spectroscopic data for NH$_{2}$CHO and its $^{13}$C isotopolog (NH$_{2}$$^{13}$CHO) were obtained from 
CDMS\footnote{\url{https://cdms.astro.uni-koeln.de}} \citep[the Cologne Database for Molecular Spectroscopy,] []{Mul01,Mul05} and \citet{Kry09}, \citet{Mot12}, while data for CH$_{3}$CONH$_{2}$ were adopted from \citet{lIy04}. The spectral modeling was performed using GILDAS\footnote{\url{http://www.iram.fr/IRAMFR/GILDAS}} (Grenoble Image and Line Data Analysis Software) to simulate the observed emission, with five parameters: source size, line width, velocity offset, rotational temperature, and column density. Of these, only the rotational temperature and column density were treated as free parameters. Source sizes were based on deconvolved continuum angular sizes; line widths were derived via Gaussian fitting; velocity offsets were calibrated using the CH$_{3}$OH line at 218440.063 MHz. Transitions heavily blended with intense lines from other species were excluded from the fitting. For partially blended transitions, multi-component Gaussian fitting was applied, and the contribution from overlapping species was estimated and subtracted using the best-fit synthetic spectra based on local thermodynamic equilibrium (LTE) modeling.

Although line intensities peak at the continuum center, strong absorption and line blending complicate the analysis. A detailed line-by-line inspection revealed that most CH$_{3}$CONH$_{2}$ transitions are blended with other species near the continuum peak, while NH$_{2}$CHO transitions suffer from significant absorption against bright continuum emission. To improve line identification in such cases, spectra were extracted from offset positions (as shown in Fig. \ref{fig:fig. A1} in the Appendix), where most of the lines exhibit Gaussian profiles and are relatively bright compared to other positions.

CH$_{3}$CONH$_{2}$ was considered confidently detected in ten HMCs (see Table \ref{table:table 2}), each showing at least five unblended transitions with signal-to-noise ratios (S/N) $\geq$ 3$\sigma$ that match the modeled rest frequencies and intensities. This criterion minimizes contamination from noise or line blending and ensures robust molecular identification. This constitutes the largest known sample of CH$_{3}$CONH$_{2}$ detections in HMCs to date and forms the basis for our subsequent analysis. Fig. \ref{fig:fig. 1} presents representative spectra of CH$_{3}$CONH$_{2}$, NH$_{2}$CHO, and NH$_{2}$$^{13}$CHO in I17175 MM2, highlighting clearly resolved or slightly blended transitions. Spectra for the other nine sources with CH$_{3}$CONH$_{2}$ detections are provided in Fig. \ref{fig:fig. B1}. Tables C.1--C.10 list all detected lines for each amide in ten HMCs.

Although several other HMCs also exhibit emission features potentially attributable to CH$_{3}$CONH$_{2}$, it is insufficient to claim a secure detection of this species in these sources, as two or less unblended transitions are observed. Thus, the focus of this study is solely on the secure detection of CH$_{3}$CONH$_{2}$. Sources that lack conclusive evidence for the existence of this species will be discussed in Appendix \ref{app:D}.

\begin{figure*}[!htp]
	\centering
    \includegraphics[width=0.9\textwidth]{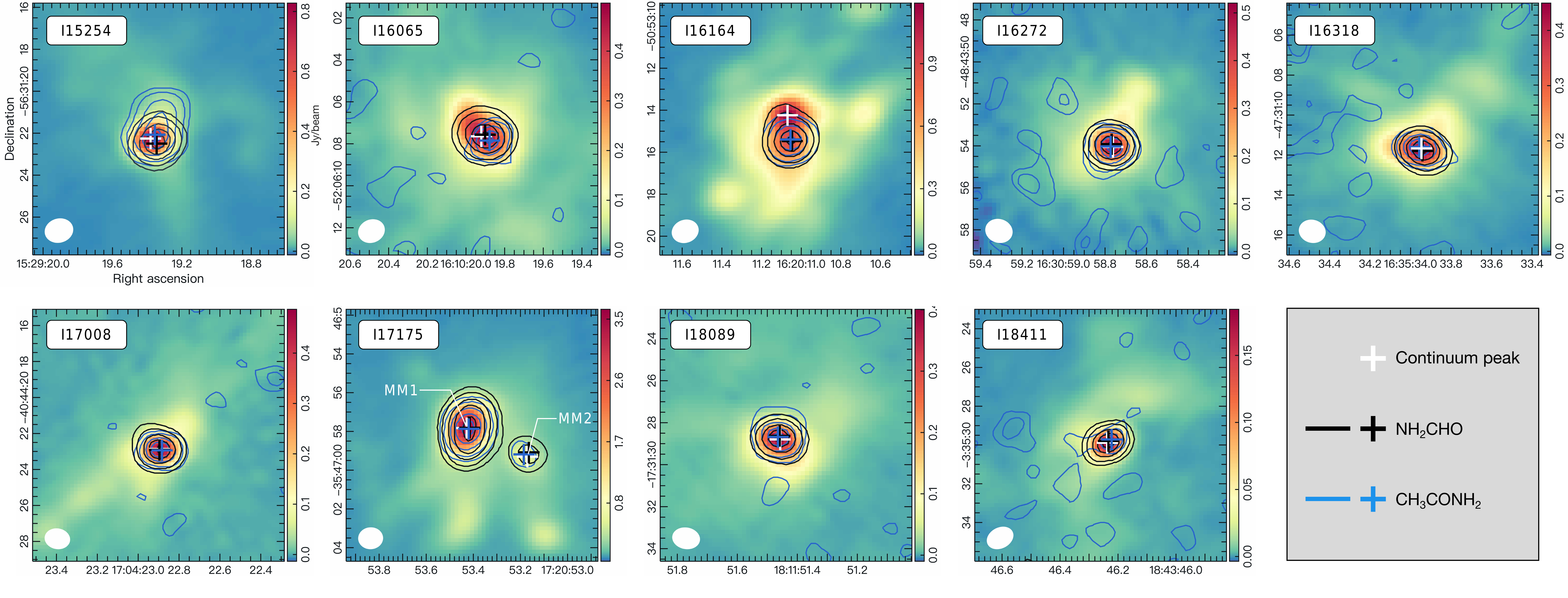}
	\caption{Integrated emission maps of NH$_{2}$CHO (black) and CH$_{3}$CONH$_{2}$ (blue), overlaid with 1.3 mm continuum emission. Contour levels are 10\%, 20\%, 40\%, 60\%, and 80\% of the peak values. The solid ellipse in the bottom left corner indicates the synthesized beam for the continuum.
	\label{fig:fig. 3}} 
\end{figure*}

\subsection{Molecular Parameter Calculations} \label{sec:sec3.2}

Column densities and excitation temperatures were derived using rotational diagrams for molecules with at least three unblended transitions spanning a broad energy range (Fig. \ref{fig:fig. 2}). This method was applied to nine sources with CH$_{3}$CONH$_{2}$ detections (excluding I18089) and to one source (I18089) with sufficient NH$_{2}$$^{13}$CHO transitions. 

For CH$_{3}$CONH$_{2}$ in I18089, the rotational diagram was not employed due to the limited energy range (117.7-136.6 K) of five detected transitions, which precludes a reliable fit. Instead, its column density was estimated from a single unblended line, assuming the excitation temperature derived from NH$_{2}$$^{13}$CHO in the same source. Similarly, for sources with fewer than three clean NH$_{2}$$^{13}$CHO transitions, column densities were estimated using individual lines and the excitation temperature obtained from CH$_{3}$CONH$_{2}$. Column densities (N) were calculated under the assumptions of LTE and optically thin emission, following the standard rotational diagram formalism \citep{Li22}:
\begin{equation}
N_{{\rm t}} = \frac{8\pi k\nu^2}{hc^3 A_{\rm ul}} \frac{Q}{g_{\rm u}} e^{E_{\rm u} / k T_{\text{ex}}} \int T_{\rm mb} \, dv
\label{equ:equ1},
\end{equation}
where $k$ is the Boltzmann constant, $\nu$ the transition frequency, $h$ the Planck constant, $c$ the speed of light, $A_{\rm ul}$ the Einstein emission coefficient, $g_{\rm u}$ the upper level degeneracy, and $E_{\rm u}$ the upper-level energy. Due to optical thickness of NH$_{2}$CHO lines, their column densities were instead derived from the optically thin NH$_{2}$$^{13}$CHO isotopolog, scaled by the appropriate $^{12}$C/$^{13}$C ratio for each source. These ratios were calculated following the equation described by \citet{Yan23} and are tabulated in Column 8 of Table \ref{table:table 1}. While this method effectively avoids opacity-related biases, several factors may introduce uncertainties into the observed $^{12}$C/$^{13}$C ratios. As noted by \citet{Yan23}, these include distance effects, beam size variations, excitation temperature mismatches, isotope-selective photodissociation, and chemical fractionation. Although the first three effects are likely minor, the latter two may be significant in regions with strong UV fields or non-equilibrium chemistry. For example, isotope-selective photodissociation may enhance the $^{12}$C/$^{13}$C ratio in photon-dominated regions (PDRs), whereas chemical fractionation could either increase or decrease the abundance of $^{13}$C-bearing species depending on formation pathways.

Table \ref{table:table 2} summarizes the derived column densities and excitation temperatures, along with abundances relative to CH$_{3}$OH. The CH$_{3}$OH column densities were adopted from \citet{Qin22}. Abundances relative to H$_{2}$ are not reported due to uncertainties in dust opacity. CH$_{3}$CONH$_{2}$ column densities range from $(2.5\pm0.9)\times10^{14}$ to $(1.5\pm0.6)\times10^{16}$ cm$^{-2}$, with excitation temperatures between 110 and 178 K. NH$_{2}$CHO exhibits consistently higher column densities, ranging from $(1.1\pm0.1)\times10^{15}$ to $(6.9\pm0.4)\times10^{16}$ cm$^{-2}$, typically 3–9 times those of CH$_{3}$CONH$_{2}$.

\subsection{Spatial distribution} \label{sec:sec3.3}

The spatial distributions of NH$_{2}$CHO and CH$_{3}$CONH$_{2}$ were analyzed using the Cube Analysis and Rendering Tool for Astronomy \citep[CARTA;] []{Com21} to investigate their morphology and potential chemical link. Fig. \ref{fig:fig. 3} presents the integrated intensity (moment-0) maps of both molecules overlaid on the 1.3 mm continuum emission. All selected transitions were individually checked to avoid contamination from line blending. Overall, amide emissions appear compact and generally coincide with the continuum peaks across all sources, except for I16164, where spatial offsets are observed. This deviation is consistent with the presence of an ultracompact H$_{\rm II}$ region reported by \citep{Zha23} and similar offsets for other COMs reported in \citet{Qin22}. In all HMCs, CH$_{3}$CONH$_{2}$ and NH$_{2}$CHO show similar spatial distributions, supporting their chemical linkage.

\section{Discussion} \label{sec:sec4}

\subsection{Detection and Distribution of CH$_{3}$CONH$_{2}$} \label{sec:sec4.1}

\begin{figure}[!htp]
	\centering
    \includegraphics[width=0.42\textwidth]{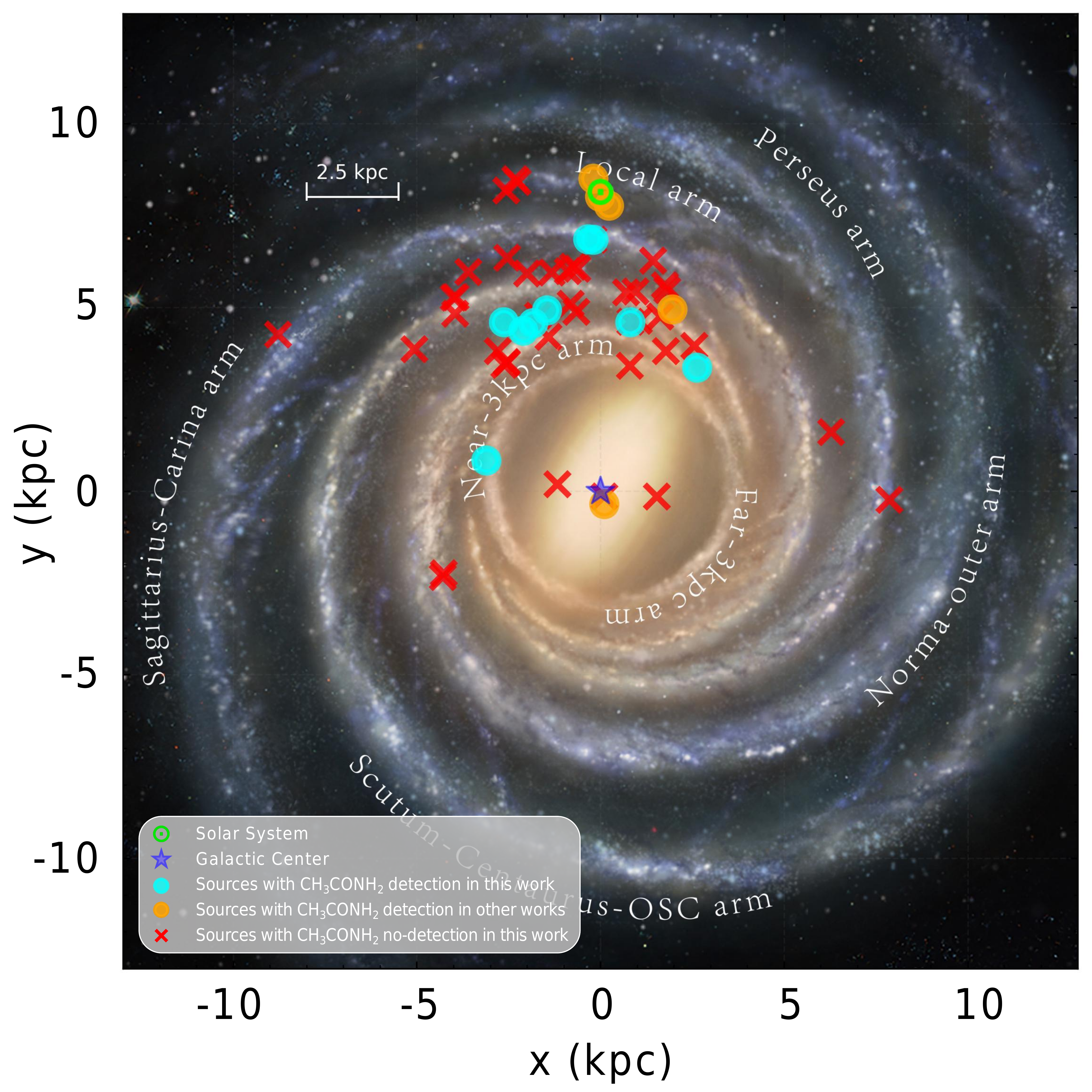}
	\caption{Spatial distribution of CH$_{3}$CONH$_{2}$-detected sources projected onto a schematic top-down view of the Milky Way. Cyan symbols represent sources with CH$_{3}$CONH$_{2}$ detections in this work, red symbols indicate sources where CH$_{3}$CONH$_{2}$ was not detected, and orange symbols denote previously reported detections from other studies. Notably, IRAS 17175 MM1/MM2 (this work) and NGC 6334I MM1/MM2 \citep{Lig20}  refer to the same HMCs and are therefore shown exclusively with cyan symbols.
	\label{fig:fig. 4}} 
\end{figure}

CH$_{3}$CONH$_{2}$, a key interstellar amide following NH$_{2}$CHO, has previously been detected in only a few individual sources. Based on unblended emission features, we report its detection in ten HMCs from the QUARKS sample, significantly increasing the number of known CH$_{3}$CONH$_{2}$ emitters in the ISM (Fig. \ref{fig:fig. 4}). This findings suggests that complex amides may be more widespread in star-forming regions than previously recognized, providing new observational constraints on their potential role in prebiotic chemistry.

\begin{figure}[!htp]
	\centering
    \includegraphics[width=0.48\textwidth]{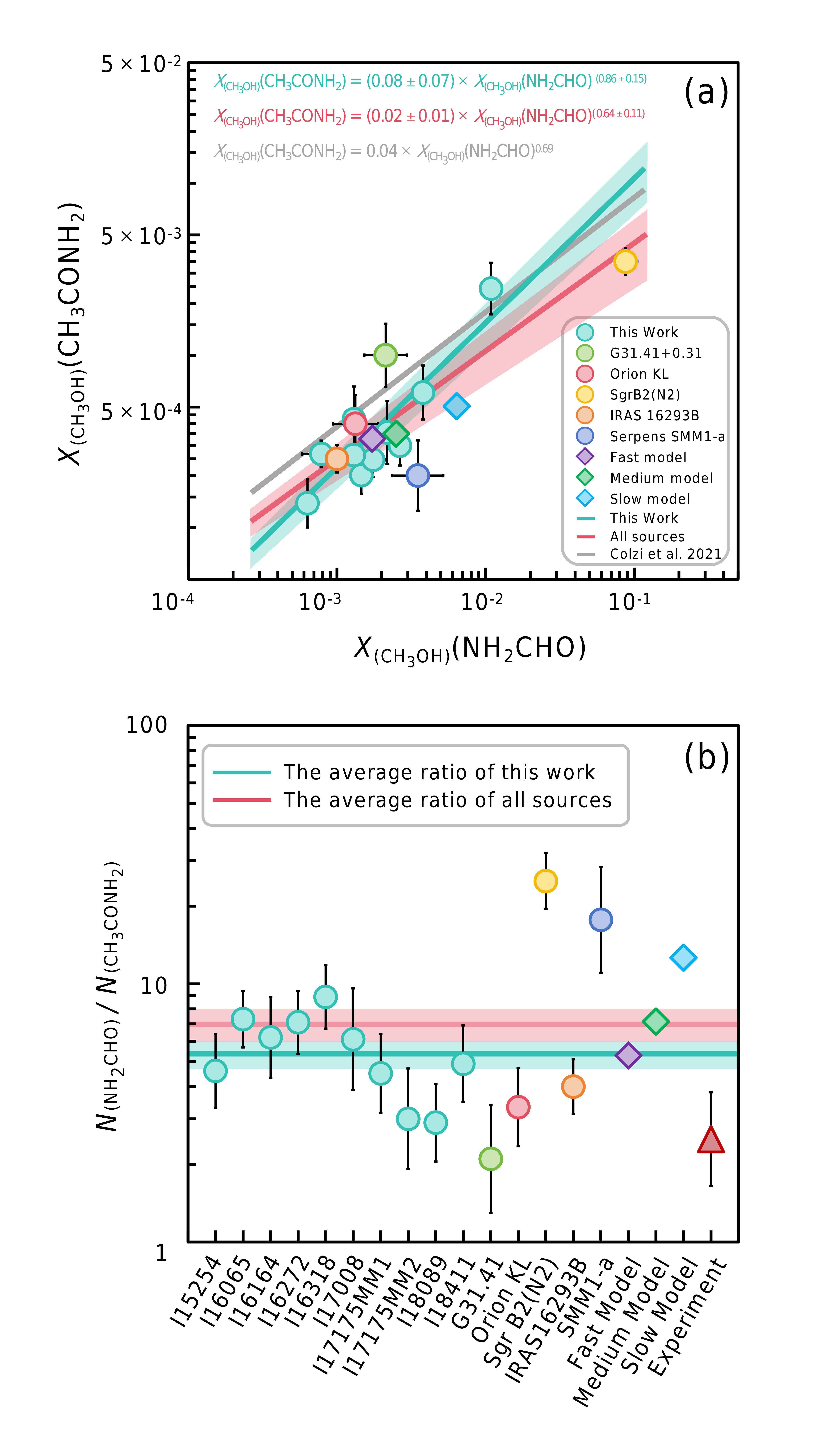}
	\caption{a) $X_{\rm CH_{3}OH}$(CH$_{3}$CONH$_{2}$) as a function of $X_{\rm CH_{3}OH}$(NH$_{2}$CHO). The cyan line and shaded region denote the power-law fit and its uncertainty derived in this work; the red line and shaded region represent the fit and its uncertainty obtained by combining our sources with previous studies; and the gray line shows the fit from \citet{Col21}. Symbols and colors correspond to different sources, as shown in the legend. b) NH$_{2}$CHO/CH$_{3}$CONH$_{2}$ ratios. Ratios for this work are indicated in cyan, while those from previous studies are shown in different colors. The cyan line and shaded region represent the mean ratio and its uncertainty for our data, and the red solid line and shaded region represent the combined sample ratio.
	\label{fig:fig. 5}} 
\end{figure}

\subsection{Chemical links between NH$_{2}$CHO and CH$_{3}$CONH$_{2}$} \label{sec:sec4.2}

To explore potential chemical connections between CH$_{3}$CONH$_{2}$ and NH$_{2}$CHO, their abundances were normalized to CH$_{3}$OH, a standard reference for COMs in the ISM. For our sources, CH$_{3}$OH column densities were adopted from \citet{Qin22}. For literature sources, values were adopted from \citealt{Cer16}, \citealt{Bog18}, \citealt{Jor18}, \citealt{Bon19}, \citealt{Lig22}, and \citealt{Min23}, assuming a 20\% uncertainty where not specified.

For comparison, we incorporated observational data from both high- and low-mass sources including Sgr B2(N2) \citep{Bel17}, G31.41+0.31 \citep{Col21}, Orion KL \citep{Cer16}, IRAS 16293-2422B \citep{Lig18}, NGC6334I MM1/MM2 \citep{Lig20}, and Serpens SMM1-a \citep{Lig22}, along with relevant chemical models \citep{Gar22} and laboratory experimental result \citep{Lig18}. While NGC6334I MM1/MM2 \citep{Lig20} correspond to the same physical sources as I17175 MM1/MM2 in our sample, the CH$_{3}$CONH$_{2}$ column densities derived in this work are lower,  likely due to more offset spectral extraction positions. Despite this, the NH$_{2}$CHO/CH$_{3}$CONH$_{2}$ abundance ratios agree well between the two studies (4.5 for MM1 and 3.0 for MM2 in our work vs. 12.3 for MM1 and 3.1 for MM2 in \citet{Lig20}), supporting the reliability of our detections. Therefore, values from NGC 6334I were excluded from further statistical analysis to avoid duplication.

As shown in Fig. \ref{fig:fig. 5}a, CH$_{3}$CONH$_{2}$ and NH$_{2}$CHO abundances exhibit a strong positive correlation, with a best-fit power law:
\begin{equation}
X_{\rm {CH_3OH}}(\rm {CH_3CONH_2}) = (0.08 \pm 0.07) \times X_{\rm {CH_3OH}}({\rm NH_2CHO})^{(0.86 \pm 0.15)}
\label{equ:equ2}.
\end{equation}
This trend indicates a potential chemically linked for the two species across diverse astrophysical environments.

The NH$_{2}$CHO/CH$_{3}$CONH$_{2}$ ratios in our sample range from 2.9 to 8.9, with an average of $\sim$5.6 and a standard deviation of $\pm$1.8, exhibiting relatively small variation across sources (Fig. \ref{fig:fig. 4}b). These values are broadly consistent with those reported for Orion KL and IRAS 16293-2422B, although they lie near the lower end of our observed range. G31.41+0.31, with a ratio of $\sim$2, is significantly lower than our sample average ($\sim$5.6), while higher ratios have been reported for Sgr B2(N2) \citep[$\sim$25;][]{Bel17} and SMM1-a \citep[$\sim$18;][]{Lig22}, exceeding average of this work by factors of $\sim$4.5, and$\sim$3.2, respectively. The observed discrepancies may arise from multiple factors. While uncertainties in excitation temperatures significantly affect the derived column densities, other contributors include optical depth, line blending, spatial structure, and assumptions of LTE. In this work, potential limitations involve the use of $^{13}$C isotopologs of NH$_{2}$CHO to correct for optical depth effects in its emission lines. Additionally, the adopted assumption of uniform source-filling emission for CH$_{3}$CONH$_{2}$ may lead to systematic uncertainties. Nevertheless, it is noteworthy that even with these discrepancies, the reported ratios from these works still fall within the same order of magnitude as our study. Given the sensitivity of COM abundances to both temperature and evolutionary stage \citep{Gar13, Gar22}, the observed stability suggests that NH$_{2}$CHO and CH$_{3}$CONH$_{2}$ likely form under similar physical and chemical conditions\citep{Que18, Bel20}.

\subsection{Formation Pathways of CH$_{3}$CONH$_{2}$} \label{sec:sec4.3}

Several formation pathways have been proposed for CH$_{3}$CONH$_{2}$. \citet{Hol06} suggested a gas-phase route involving the addition of a CH$_{2}$ radical to NH$_{2}$CHO. However, theoretical studies by \citet{Qua07} indicate that this reaction requires a spin-flip transition of CH$_{2}$ from the triplet to singlet state, with an energy barrier exceeding 1000 K—rendering it unlikely under typical interstellar conditions. Other proposed gas-phase reactions, such as radiative association between NH$_{2}$CHO and CH$^{+}_{3}$ \citep{Qua07}, predict CH$_{3}$CONH$_{2}$ abundances ($\sim$10$^{-15}$) far below observed levels.

In contrast, grain-surface chemistry offers a more viable formation mechanism. \citet{Aga85} proposed that CH$_{3}$CONH$_{2}$ can form via the reaction between the CH$_{3}$ radical and the NH$_{2}$CO intermediate—a key species linking NH$_{2}$CHO and CH$_{3}$CONH$_{2}$. The NH$_{2}$CO radical may arise through hydrogen abstraction from NH$_{2}$CHO, radical addition of NH$_{2}$ and CO, or hydrogenation of HNCO \citep{Bel17, Bel19, Gar22}. \citet{Bel17} predicted an NH$_{2}$CHO/CH$_{3}$CONH$_{2}$ ratio of $\sim$15 assuming only the hydrogen abstraction route. In contrast, \citet{Gar22} considered multiple grain-surface formation pathways for CH$_{3}$CONH$_{2}$, including: (i) CH$_{3}$ + NH$_{2}$CO → CH$_{3}$CONH$_{2}$, (ii) H-abstraction from NH$_{2}$CHO forming NH$_{2}$CO, followed by CH$_{3}$ addition, and (iii) hydrogenation of HNCO leading to NH$_{2}$CO intermediates. These mechanisms all involve NH$_{2}$CO as a key radical intermediate, thereby establishing a direct chemical connection between NH$_{2}$CHO and CH$_{3}$CONH$_{2}$. Our observed abundance ratios (2.9–8.9) align with Garrod et al.’s predicted range (5–13), providing strong support for a linked formation pathway through grain-surface chemistry.

The strong abundance correlation between NH$_{2}$CHO and CH$_{3}$CONH$_{2}$ may also reflect their co-formation on dust grains, followed by thermal desorption under similar physical conditions—paralleling trends observed for HNCO and NH$_{2}$CHO \citep{Que18}. Laboratory data further support this view: \citet{Lig18} measured similar desorption temperatures for CH$_{3}$CONH$_{2}$ (219 K) and NH$_{2}$CHO (210 K), while ice mantle simulations yielded a desorption ratio of NH$_{2}$CHO/CH$_{3}$CONH$_{2}$ $\sim$2.5$^{+1.3}_{-1.2}$, comparable to our average observed value of 5.6. 

Taken together, these findings strongly favor a formation scenario in which CH$_{3}$CONH$_{2}$ is synthesized via grain-surface chemistry closely linked to NH$_{2}$CHO, followed by co-desorption in star-forming regions. Nonetheless, additional laboratory, observational, and modeling efforts are required to fully constrain the dominant formation routes.

\section{Conclusions} \label{sec:sec5}
We present the first systematic survey of acetamide (CH$_{3}$CONH$_{2}$), a key interstellar amide, in hot molecular cores (HMCs) using ALMA-QUARKS data. The joint detections of formamide (NH$_{2}$CHO) enable a comparative analysis of their abundance correlation and formation chemistry. Our main findings are as follows:

   \begin{enumerate}
      \item Both NH$_{2}$CHO and CH$_{3}$CONH$_{2}$ were detected in ten HMCs, increasing the number of known CH$_{3}$CONH$_{2}$ sources to 17 and providing the most extensive sample to date for this molecule.   
         
      \item CH$_{3}$CONH$_{2}$ column densities, derived via rotational diagram analysis, range from $(2.5\pm0.9)\times10^{14}$ to $(1.5\pm0.6)\times10^{16}$ cm$^{-2}$. NH$_{2}$CHO column densities, estimated from optically thin $^{13}$C isotopologs, range from $(1.1\pm0.1)\times10^{15}$ to $(6.9\pm0.4)\times10^{16}$ cm$^{-2}$ --- typically 3--9 times those of CH$_{3}$CONH$_{2}$.     
      
      \item The NH$_{2}$CHO/CH$_{3}$CONH$_{2}$ ratio remains nearly constant ($\sim$5.6 on average) across all detected sources. A strong correlation between the two species follows a power-law relation, supporting a chemically linked formation pathway, likely dominated by grain-surface reactions.\\
   \end{enumerate}  

These results provide compelling evidence for the presence of complex amide molecules in star-forming regions. The inferred chemical association between NH$_{2}$CHO and CH$_{3}$CONH$_{2}$ highlights the potential role of surface chemistry in assembling peptide-relevant structures prior to planetary formation. The eventual incorporation of such molecules into nascent planetary systems -- via cometary or meteoritic delivery -- may have contributed essential building blocks for prebiotic chemistry on early Earth and potentially elsewhere. Continued laboratory, observational, and modeling efforts are required to further constrain the dominant formation routes and the broader astrochemical role of interstellar amides.

\section{Data availability} \label{sec:sec6}
The derived data underlying this article are available in the article and in its online supplementary material on \href{https://doi.org/10.5281/zenodo.16673390}{zenodo}.

\begin{acknowledgements}

This work makes use of the following ALMA data: ADS/JAO.ALMA\#2021.1.00095.S. ALMA is a partnership of ESO (representing its member states), NSF (USA), and NINS (Japan), together with NRC (Canada), MOST and ASIAA (Taiwan, China), and KASI (Republic of Korea), in cooperation with the Republic of Chile. The Joint ALMA Observatory is operated by ESO, AUI/NRAO, and NAOJ. We sincerely thank Prof. Shengli Qin and Dongting Yang (Yunnan University) for their assistance with the data processing. This work was supported by the Fundamental Research Funds for the Central Universities (Grant Nos. 2024CDJGF-025 and 2023CDJXY-045 ), Chongqing Municipal Natural Science Foundation General Program (Grant No. cstc2021jcyj-msxmX0867), National Natural Science Foundation of China (Grant No. 12103010), and Strategic Priority Research Program of the Chinese Academy of Sciences (Grant No. XDB0800303).
      
\end{acknowledgements}

\bibliographystyle{aa} 
\bibliography{aa56073-25.bib} 

\begin{appendix}

\onecolumn

\section{Spectral Extraction positions.}
\label{app:A}

\begin{figure*}[!htbp]
    \centering
    \includegraphics[width=1\textwidth]{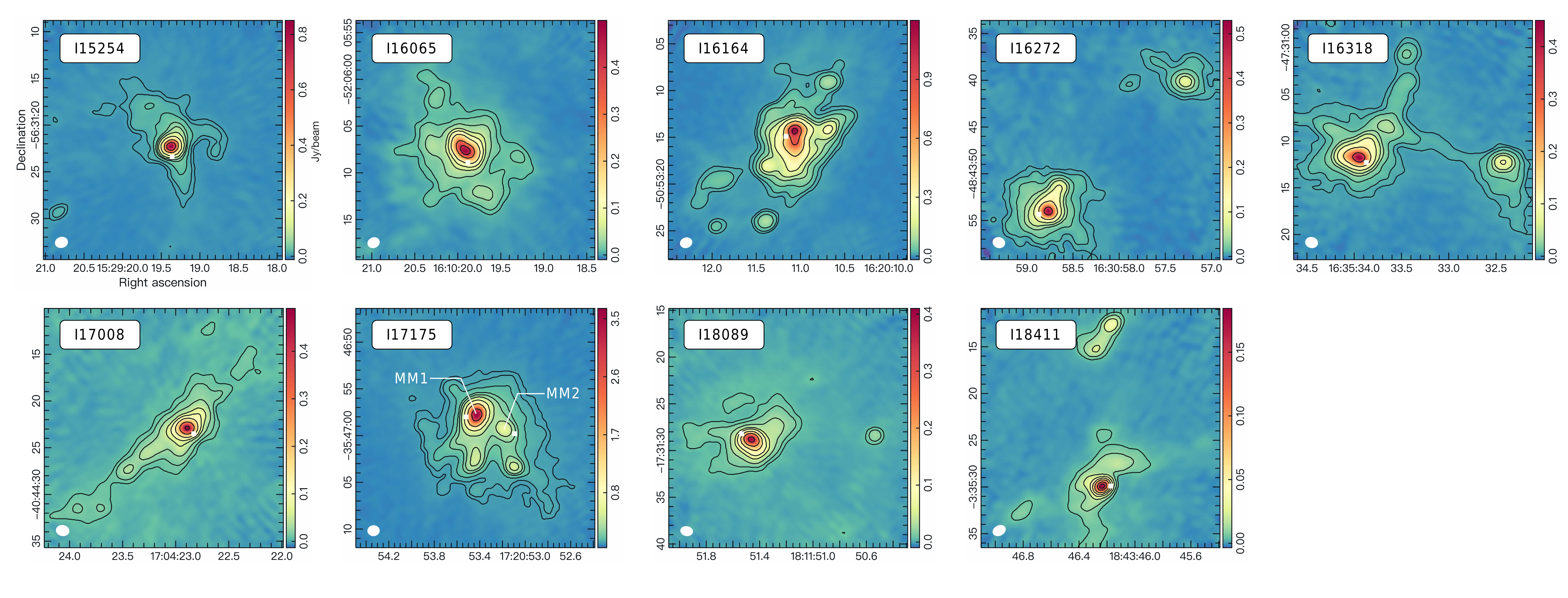}
    \caption{1.3 mm continuum image of I15254, I16065, I16164, I16272, I16318, I17008, I17175, I18089, and I18411 obtained with ALMA band 6. The background shows the continuum emission of each source. Contour levels are drawn at (3, 5, 10, 20, 30, 50, 100, 200) $\times \sigma$, where $\sigma$ is the rms noise level. The 1$\sigma$ values of I15254, I16065, I16164, I16272, I16318, I17008, I17175, I18089, and I18411 are 1.92, 2.75, 4.13, 1.75, 1.51, 2.80, 6.90, 2.60, 2.80 mJy beam$^{-1}$, respectively. The synthetic beam for continuum is indicated in the bottom left corner by the ellipse. The spectral extraction positions are indicated with a white point.
    \label{fig:fig. A1}} 
\end{figure*}

To minimize the effects of absorption and line blending, we extracted spectra at offset positions from the continuum peaks for further analysis. The extraction positions for each source are shown in Fig. \ref{fig:fig. A1}, where they are marked by white points.

\section{The identified spectral lines of amide molecules.}
\label{app:B}

\begin{figure*}[!htbp]
    \centering
    \includegraphics[width=1\textwidth]{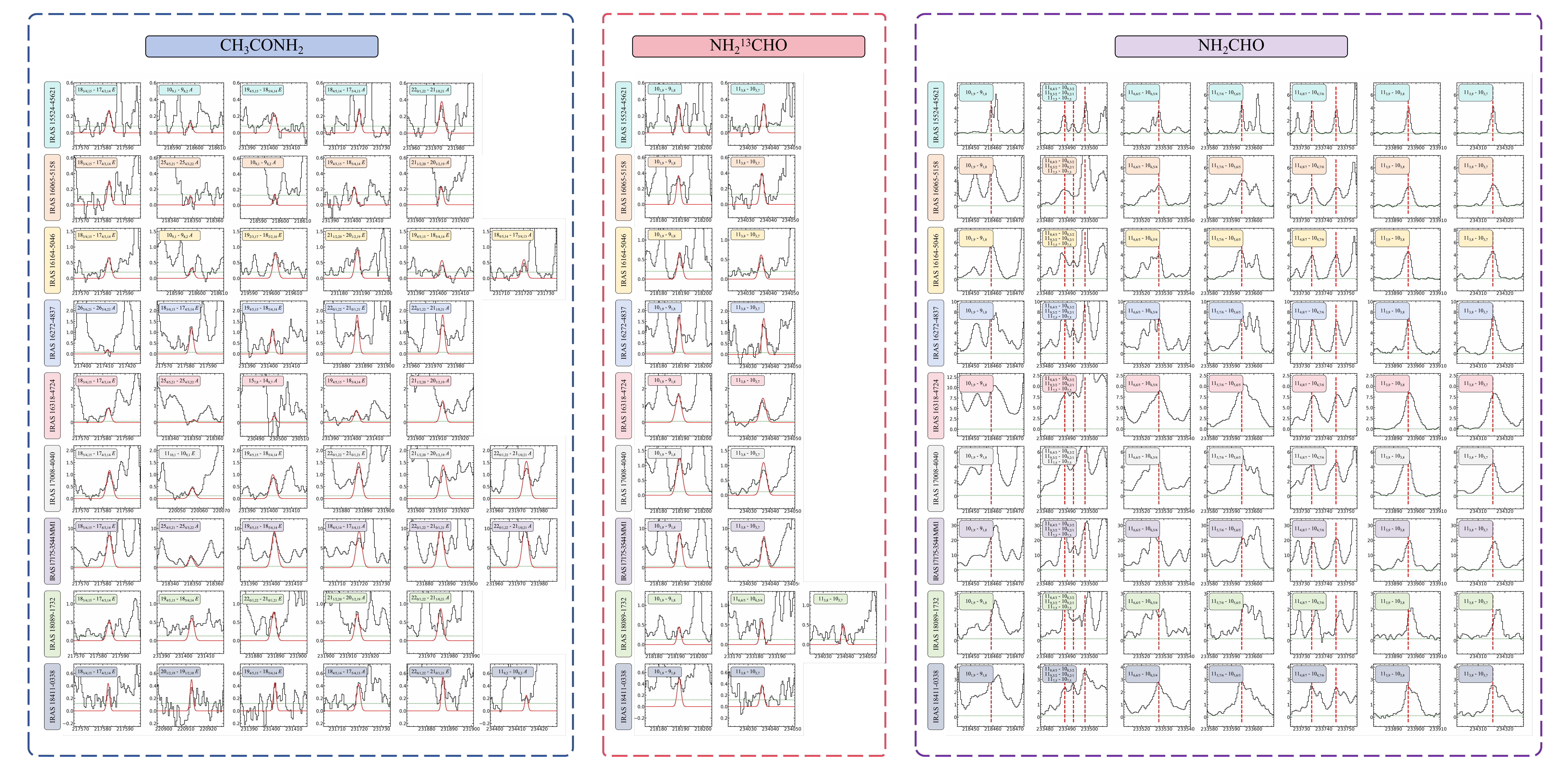}
    \caption{Observed (black) and modeled (red) spectra of CH$_{3}$CONH$_{2}$, NH$_{2}$CHO, and NH$_{2}$$^{13}$CHO in nine HMCs. Clearly resolved or slightly blended transitions are shown. Green dashed lines mark the 3$\sigma$ noise level. No modeled spectrum is shown for NH$_{2}$CHO due to optical thickness. Red dashed lines indicate the rest frequencies of NH$_{2}$CHO transitions. The y-axis shows the line intensity in Kelvin, and the x-axis shows the frequency in MHz.
    \label{fig:fig. B1}} 
\end{figure*}

The emissions of CH$_{3}$CONH$_{2}$, NH$_{2}$CHO, and NH$_{2}$$^{13}$CHO are detected in ten sources. Fig. \ref{fig:fig. B1} show the detected transitions of these amide molecules, which are not covered in the Fig. \ref{fig:fig. 1}, for all hot cores.

\section{Summary of detected molecular lines.}
\label{app:C}

This appendix presents the statistics of detected lines in our observations. Tables C.1--C.10, only available at the \href{https://doi.org/10.5281/zenodo.16673390}{zenodo}, compile all detected line information for each amide species across the ten HMCs. These data provide essential constraints for the derivation of column densities and excitation temperatures.

\section{The upper limits of column density for CH$_3$CONH$_2$ in undetected sources.}
\label{app:D}

\begin{figure*}[!htbp]
    \centering
    \includegraphics[width=1\textwidth]{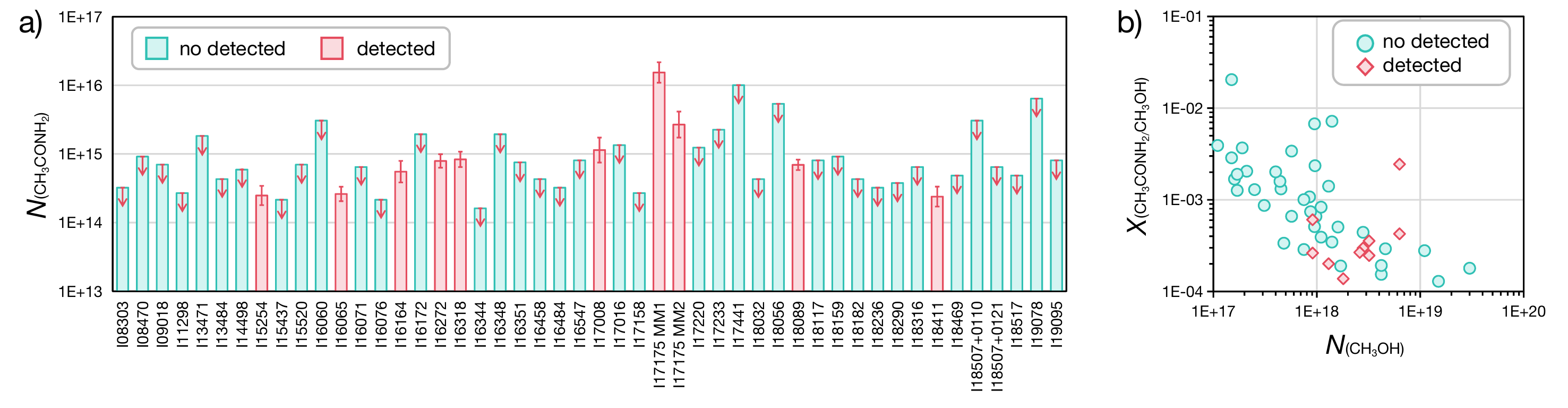}
    \caption{(a) Derived column densities (10 sources with CH$_3$CONH$_2$ detections, indicated in red) or upper limits (42 sources without CH$_3$CONH$_2$ detections, indicated in cyan) among all sources. (b) The relationship between the CH$_3$CONH$_2$/CH$_3$OH and CH$_3$OH column densities.
    \label{fig:fig. D1}} 
\end{figure*}

For the 42 sources where CH$_3$CONH$_2$ was not detected, 3$\sigma$ upper limits for its column density has been derived. Fig \ref{fig:fig. D1}(a) compares these upper limits with the derived column densities in the 10 detections. The upper limits generally fall within an order of magnitude of the detected values, suggesting that sensitivity limitations may account for many of the non-detections, rather than a true absence of CH$_3$CONH$_2$. To further explore this, the relationship between CH$_3$CONH$_2$/CH$_3$OH ratios and CH$_3$OH column densities have been examined (Fig \ref{fig:fig. D1}(b)). All ten CH$_3$CONH$_2$-detected sources exhibit higher CH$_3$OH column densities, supporting the idea that CH$_3$CONH$_2$ is more likely to be detected in chemically rich environments. Nonetheless, 6 non-detected sources with CH$_3$OH column densities > $4\times10^{18} {\rm cm^{-2}}$, show no CH$_3$CONH$_2$ detection, suggesting that  CH$_3$OH abundance alone is not a sufficient predictor. These results emphasize the need for higher-sensitivity observations to better constrain the prevalence and formation conditions of interstellar acetamide.

\end{appendix}

\end{document}